\documentclass[twoside,9pt]{article}

\usepackage[super,sort&compress,comma]{natbib} 

\usepackage{times,mathptm}
\usepackage{sectsty}
\usepackage{balance} 
\usepackage[text={11.3cm,20.4cm},centering]{geometry} 

\usepackage{graphicx}
\usepackage{lastpage}
\usepackage[format=plain,justification=raggedright,singlelinecheck=false,font=small,labelfont=bf,labelsep=space]{caption} 
\usepackage{fancyhdr}

\begin{document}

\thispagestyle{plain}
\fancypagestyle{plain}{
%\fancyhead[L]{\includegraphics[height=8pt]{headers/LH}}
%\fancyhead[C]{\hspace{-1cm}\includegraphics[height=15pt]{headers/CH}}
%\fancyhead[R]{\includegraphics[height=10pt]{headers/RH}\vspace{-0.2cm}}
\renewcommand{\headrulewidth}{1pt}}
\renewcommand{\thefootnote}{\fnsymbol{footnote}}
\renewcommand\footnoterule{\vspace*{1pt}% 
\hrule width 11.3cm height 0.4pt \vspace*{5pt}} 
\setcounter{secnumdepth}{5}

\makeatletter 
\renewcommand{\fnum@figure}{\textbf{Fig.~\thefigure~~}}
\def\subsubsection{\@startsection{subsubsection}{3}{10pt}{-1.25ex plus -1ex minus -.1ex}{0ex plus 0ex}{\normalsize\bf}} 
\def\paragraph{\@startsection{paragraph}{4}{10pt}{-1.25ex plus -1ex minus -.1ex}{0ex plus 0ex}{\normalsize\textit}} 
\renewcommand\@biblabel[1]{#1}            
\renewcommand\@makefntext[1]% 
{\noindent\makebox[0pt][r]{\@thefnmark\,}#1}
\makeatother 
\sectionfont{\large}
\subsectionfont{\normalsize} 

%\fancyfoot{}
%\fancyfoot[LO,RE]{\vspace{-7pt}\includegraphics[height=8pt]{headers/LF}}
%\fancyfoot[CO]{\vspace{-7pt}\hspace{5.9cm}\includegraphics[height=7pt]{headers/RF}}
%\fancyfoot[CE]{\vspace{-6.6pt}\hspace{-7.2cm}\includegraphics[height=7pt]{headers/RF}}
\fancyfoot[RO]{\scriptsize{\sffamily{1--\pageref{LastPage} ~\textbar  \hspace{2pt}\thepage}}}
\fancyfoot[LE]{\scriptsize{\sffamily{\thepage~\textbar\hspace{3.3cm} 1--\pageref{LastPage}}}}
\fancyhead{}
\renewcommand{\headrulewidth}{1pt} 
\renewcommand{\footrulewidth}{1pt}
\setlength{\arrayrulewidth}{1pt}
\setlength{\columnsep}{6.5mm}
\setlength\bibsep{1pt}

\noindent\LARGE{\textbf{Reassignment of magic numbers for icosahedral Au clusters: 310, 564, 928 and 1426}}
\vspace{0.6cm}

\noindent\large{\textbf{Jan Kloppenburg,\textit{$^{a}$} 
Andreas Pedersen,\textit{$^{b}$} Kari Laasonen,\textit{$^{c}$} 
Miguel A. Caro\textit{$^{a}$}
and Hannes J\'onsson$^{\ast}$\textit{$^{b,d}$}}

\vspace{0.6cm}

\noindent\normalsize{
Icosahedral Au clusters with three and four shells of atoms are found to deviate significantly from the commonly assumed Mackay structures. By introducing additional atoms in the surface shell and creating a vacancy in the center of the cluster, the calculated energy per atom can be lowered significantly, according to several different descriptions of the interatomic interaction. Analogous icosahedral structures with five and six shells of atoms are generated using the same structural motifs and are similarly found to be more stable than Mackay icosahedra. The lowest energy per atom is obtained with clusters containing 310, 564 , 928 and 1426 atoms, as compared with the commonly assumed magic numbers of 309, 561, 923 and 1415. Some of the vertices in the optimized clusters have a hexagonal ring of atoms, rather than a pentagon, with the vertex atom missing. An inner shell atom in some cases moves outwards by more than an \AA ngstr\"om into the surface shell at the vertex site. This feature, as well as the wide distribution of nearest-neighbor distances in the surface layer, can strongly influence the catalytic properties of icosahedral clusters. The structural optimization is initially carried out using the GOUST method with atomic forces estimated with the EMT empirical potential function, but the atomic coordinates are then refined by minimization using electron density functional theory (DFT) or Gaussian approximation potential (GAP). A single energy barrier is found to separate the Mackay icosahedron from a lower energy structure where a string of atoms moves outwards in a concerted manner from the center so as to create a central vacancy while placing an additional atom in the surface shell.
}

\renewcommand*\rmdefault{bch}\normalfont\upshape
\rmfamily
\section*{}
\vspace{-1cm}
\footnotetext{\textit{$^{a}$~Department of Electrical Engineering and Automation, Aalto University, FIN-02150 Aalto, Finland.}}
\footnotetext{\textit{$^{b}$~Science Institute and Faculty of Physical Sciences, University of Iceland VR-III, 107 Reykjav\'{\i}k, Iceland.}} 
\footnotetext{\textit{$^{c}$~Department of Chemistry and Materials Science, Aalto University, 
%P.O. Box 16100, 
FI-00076 Aalto, Finland}}
\footnotetext{\textit{$^{d}$~Department of Applied Physics, Aalto University, FI-00076 Aalto, Finland}}

% --------------------------------------------------------------------------------------------------------------------------------------
      
\section{Introduction}

Gold nanoclusters have many intriguing properties and can be used in various applications such as 
catalysis\cite{Haruta97,Valden98,Corma13,Saint-Lager13},
plasmonics\cite{Maier01,Ghosh07} and
biosensors.\cite{Saha12}
In order to gain an understanding of their functionality, the atomic structure of the clusters needs to be identified.
A striking example of the variation of functionality with size is the remarkably facile oxidation of CO 
on Au clusters with diameter between 2 to 3 nm adsorbed on TiO$_2$.\cite{Saint-Lager13} 
Clusters outside this size range, both larger and smaller, have been found to be less active per surface site.
The ordering of the atoms in the clusters, in particular on the cluster surface, strongly influences the catalytic properties.
Low coordinated Au atoms on the surface are believed to be responsible for the catalytic activity (see, for example, ref. \cite{Brodersen11}),
but the reduced efficiency for clusters with diameter smaller than 2 nm remains a puzzle.

Small Au clusters are planar but 
clusters with 13 or more atoms have three-dimensional structure.\cite{Assa09}
Various pyramids and cages have been deduced from experimental and computational studies 
of clusters with up to a few tens of atoms.
\cite{Michaelian99,Li03,Baletto05,Pyykko08,Huang08,Wang12a}
The rich size dependence of Au clusters in this size range has been ascribed to relativistic effects.\cite{Pyykko08}

When the interaction between atoms does not depend strongly on the directionality of the chemical bonds, 
atomic clusters tend to have icosahedral shape and five-fold symmetry up to a certain cluster size.
This applies, for example, to rare gas atoms and simple metals, where the interaction energy mostly depends on 
the distance between atoms. 
Calculations of clusters where the atomic interaction is described with a simple pairwise Lennard-Jones potential function
have shown that the icosahedral structure motif dominates for clusters with up to 5000 atoms.\cite{Honeycutt87}
Eventually, for large enough clusters, the atoms order in a way that is characteristic of the crystalline phase as it becomes lower in energy
than icosahedral ordering.
Mackay presented the atomic structure of icosahedral clusters of increasing size as shells of atoms are added to the 
basic 13 atom cluster.  
\cite{Mackay62}  
The surface layer is complete and compact when the number of atoms in the cluster is 13, 55, 147, 309, 561, 923, 1415 \dots
These are often referred to as `magic numbers' since the energy per atom can then be particularly low and they can, therefore,
be expected to be more abundant when clusters of various size are generated experimentally.

Clusters with Ino-decahedral\cite{Ino69} and cuboctahedral shapes turn out to have the same magic numbers as the Mackay
icosahedra.
Marks decahedra\cite{Marks84,Marks94} involve different truncations of the twinned tetrahedra and give rise to different magic numbers. 
\cite{Doye95,Bao09}
For noble gases and simple metals, a correlation is indeed found between the abundance of clusters as a function of size and the magic numbers.
Some theoretical and experimental studies have assumed that the same magic numbers apply to Au clusters.

The electronic structure of metal clusters also favors clusters of a certain size corresponding to complete 
valence electron shells leading to extra stability for clusters 
with 8, 18, 34, 58, and 92 atoms.\cite{Knight84,Wrigge02,Larsen11} 
But, for larger clusters this effect becomes relatively less important than the packing of the atoms. 

Many studies have been devoted to Au clusters with 55 atoms, one of the traditional magic numbers.  
The optimal structure was unexpectedly predicted 
to be mostly amorphous but nevertheless chiral.\cite{Garzon96,Garzon98,Michaelian99} 
Local structure analysis has identified part of it as a distorted icosahedron.
Photoelectron spectroscopy measurements of ionic Au$_{55}^-$ clusters combined with density functional theory calculations\cite{Huang08}
as well as aberration corrected scanning transmission electron microscope (AC-STEM) measurements of Au$_{55}$ clusters
deposited on a amorphous carbon surface\cite{Wang12a}
have lent support for this structural prediction. 
The origin of the of the low symmetry of the Au$_{55}$ cluster has been ascribed to a strong contractions of the surface of the cluster, 
thereby creating vacancy islands.
As the cluster grows up to Au$_{64}$, additional atoms get introduced into the surface layer and
the energy per atom is reduced as a result.\cite{Huang08}  
This is analogous to gold crystal surfaces, such as Au(100), where the additional atoms get inserted into the surface layer in 
the so-called hex reconstruction.
\cite{vanHove81,Takeuchi91} 

As the cluster size increases, the icosahedral (Ih) structure becomes less stable with respect to the decahedral (Dh)  
and cuboctahedral structures, the latter having the same local atomic ordering as the face centered cubic crystal, exposing (111) and (100) facets at
the surface.  The (100) facets are higher in energy so a lower energy structure is generally obtained by truncating the 
cuboctahedron in such a way as to reduce the area of the (100) facets to form a truncated cuboctahedron (TCo).
It has often been assumed that the crossover from one structural type to another is a sharp function of cluster size and
that it follows the progression
Ih $\rightarrow$ Dh $\rightarrow$ TCo.\cite{Baletto02,Li08}
The determination of these crossovers has been the topic of many computational studies and various predictions have been made. 
The crossover from Ih to Dh has generally been predicted to occur for relatively small clusters, with less than 150 atoms. 
A larger variety of predictions have been made for the Dh to TCo crossover, with estimates ranging from 200 to 700 atoms\cite{Baletto02,Grochola07,Cleveland97,Uppenbrink92} or even larger than 
5000 atoms.\cite{Negreiros07,Barnard09} 

Other calculations have, however, indicated that the crossover between structural motifs is not so simple.\cite{Bao09,Rahm17,Garden18} 
While the energy per atom of Dh clusters varies smoothly with size, as the structure can be that of Ino or Marks decahedra 
or some variation on those motifs,
there are larger fluctuations for the TCo motif where the regular crystalline facets provide the highest stability.
This results in alternations of the most stable structural motif over a wide size range, up to around 4000 atoms.\cite{Garden18}
As a result, different magic numbers are obtained for Dh and TCo clusters than one would expect from the 
shell closing of the ideal structures mentioned above.

Recent advances in experimental imaging techniques, in particular the development of the aberration corrected scanning transmission electron microscope (AC-STEM), 
have led to direct observations of three-dimensional, atomic-scale structure of Au clusters.\cite{Li08,Curley07,Wang12b,Plant14,Wells15} 
The AC-STEM method has been used to study larger Au clusters deposited on a support, even as large as 1000 atoms.
While the preparation method can be tuned to promote the formation of certain structures,\cite{Plant14}
the inherent metastability of clusters and energy barriers for transitions from one structure to another make
it difficult to attribute experimentally observed isomeric populations to equilibrium populations.\cite{Wells15} 
A further complication is the possible temperature dependence of the optimal structure due to entropic effects.\cite{Marks94,Koga04}

The atomic-scale structure of Au clusters with up to around 1000 atoms has been investigated extensively in recent years by Palmer and coworkers
using AC-STEM.\cite{Li08,Plant14,Wang12a,Wang12b,Wells15} 
All three structural motifs, Ih, Dh and TCo, have been observed over this size range. 
The icosahedral structure is deduced to be the least stable of the three motifs for clusters with 923 atoms, since it has lower 
abundance\cite{Li08,Plant14} and is found to transform to one of the other motifs under beam irradiation.\cite{Wang12b,Wells15}  
Determining the relative stability of the Dh and TCo motifs has turned out to be more difficult. 
For Au$_{923}$, the Dh and TCo clusters generally retain their structural motif under irradiation suggesting that both are 
stable structures\cite{Wang12b} 
but for Au$_{561}$ the Dh isomer was found to transform to TCo.
\cite{Wells15} 
Interestingly, the populations of Dh and TCo for 561, 742 and 923 atom Au clusters were found to be nearly the same, within experimental uncertainty. 
Based on classical dynamics simulations of cluster growth where transformations between structures were not observed during the growth process,
it was concluded that the structure of a cluster becomes `locked in' at a small size after which templated growth takes place without
transformation to a more stable structure.\cite{Wells15} 
An alternative explanation can be formulated based on the close but alternating stability of the Dh and TCo structures with cluster size, mentioned above.\cite{Garden18}

Despite the large amount of research to date, the motif preference for medium size (300 to 1000 atoms) Au clusters remains an open question. Furthermore, the prediction of the most stable atomic structure for a given structural motif has turned out to be a challenge.  
We study here the atomic structure of Au nanoclusters with the icosahedral motif in this size range and search for those 
that have particularly low energy per atom. 
The global minimization of the energy with respect to atom coordinates for clusters of this size is challenging and it is hard to be
sure that the optimal solution has been found, but it is clear from the results obtained that the usual magic numbers do not apply
to Au clusters.

% --------------------------------------------------------------------------------------------------------------------------------------

\section{Methods}

The structure optimization of the clusters with three and four atom shells 
is carried out with a method referred to as global optimization using saddle traversals (GOUST).\cite{GOUST}
There, the system is brought from one local minimum on the potential energy surface to another by identifying low lying first order saddle points. 
From each local minimum, several saddle point searches are conducted and the local energy minimum on the opposite side 
of the energy ridge is found by 
minimization after a slight displacement along the eigenvector corresponding to the negative eigenvalue of the Hessian matrix.
The system is then advanced to the lowest local minimum found in the various saddle point searches carried out for this structure. 
This process is repeated for several different random initial configurations of the atoms.

The saddle points are found  
using the minimum mode following (MMF) method.\cite{Henkelman99,Pedersen11,Gutierrez17}
The EON software \cite{Chill14} and the EMT interaction potential function\cite{Jacobsen96}  
(with Rasmussen parameter values)
are used for the initial phase of the structural optimization.
Subsequently, a local minimization of the energy is carried out using density functional theory (DFT) with either the PBE\cite{Perdew96} or the PBEsol functional approximations.\cite{Perdew08} 
The PBEsol functional is known to give better estimates of the surface energy of metals,
so it is expected to be more accurate for small clusters where the surface energy is of great importance. 
The difference in energy obtained with the two functionals can be used as an indication of the accuracy of the DFT calculations.
A plane-wave basis with a 230~eV kinetic energy cutoff was used to represent the valence electrons 
and the projector augmented wave formalism\cite{Blochl94} for the inner electrons,
as implemented in the VASP software.\cite{Kresse96a,Kresse96b}
 
Further analysis was carried out using a Gaussian approximation potential (GAP)\cite{GAPa,GAPb},
fitted to an extensive data set of PBE calculations, including various types of crystal structures, surface slabs
and clusters with random disorder (which mimics thermal excitation).
GAP is a data-driven approach to interpolating the interatomic energies and forces based on Gaussian process
regression.\cite{Deringer21} 
The GAP potential used here for Au atoms\cite{Kloppenburg22} incorporates two-body and atom-centered many-body
descriptors\cite{Bartok13,Caro19} to encode the
local atomic environments. This leads to an atom-wise decomposition of the total energy which can be used
to analyze further the driving force for the structural changes in the gold clusters during optimization.
Starting with the cluster structures obtained by GOUST and EMT, the GAP energy is minimized with respect to the atomic 
coordinates and the energy per atom evaluated. 
This also allows for a comparison of the DFT and GAP total energy changes to assess the accuracy of the machine learned potential function.  
 
Guided by the results obtained for clusters with three and four shells, 
atomic structures were constructed for icosahedral clusters with five and six shells of atoms.
This involved inserting rows of additional atoms in various ways and then minimizing the energy. 
The minimization was carried out using the GAP potential as well as with DFT.

% --------------------------------------------------------------------------------------------------------------------------------------

\section{Results}

After carrying out a few tens of saddle point searches starting with the 309 atom Mackay icosahedron and small initial random displacements
of the atoms, 
a concerted displacement of 6 atoms was found where the central atom moves into the first (12 atom) shell, 
taking the place of an atom that moves from the first shell to the second (42 atom) shell and the
atom initially in the second shell moves into the third shell, thus adding a new atom to the surface of the cluster. 
This transition corresponds to a single saddle point on the energy surface. 
A similar process involving a concerted displacement of 7 atoms was found for the larger 561 atom Mackay icosahedron. 
The energy along the minimum energy path and an illustration of the displacement of the atoms of the 561 atom cluster is shown in Fig. 1.
The final state of this transition is 0.8 eV lower in energy as both the compressive strain on the central atom and
the tensile strain in the surface shell is relieved. As can be seen from the energy curve in Fig. 1, there is just a single
maximum along the minimum energy path for this multi-atom transition mechanism. 
Subsequent steps in the global optimization result in the formation of vacancy patches in the surface 
shell as surface atoms group together,  
making room for additional atoms to be incorporated into the surface shell.
Sequential addition of atoms in various places in or near the surface shell followed by structural optimization shows that
the lowest energy per atom is obtained by adding one atom to the 309 atom cluster and adding three atoms to the 561 atom cluster.
The optimal structures were further refined by local minimization using DFT or GAP, followed by analysis of the structure and 
energy per atom.

% -------------------------------------  Fig. 1 -------------------------------------------------------
\begin{figure}
\center
\includegraphics[width=.9\textwidth]{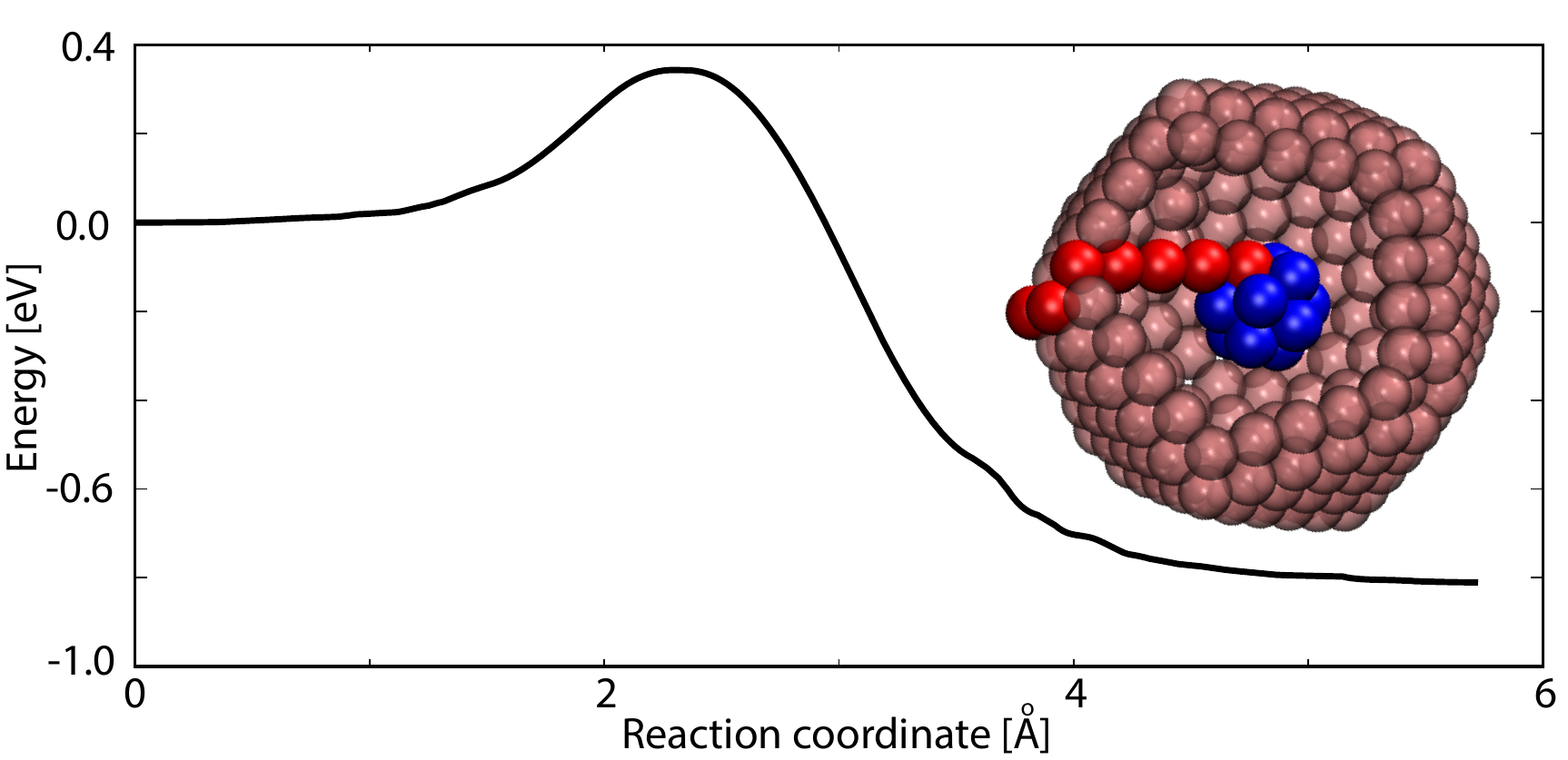}
\caption{
Energy along the minimum energy path for a transition in a 561 atom Au cluster where the initial
structure is the energy minimized Mackay icosahedron and the
final state (shown in inset) has a vacancy in the center and an extra atom in the surface shell.
The transition involves a
concerted displacement of seven atoms (shown in red) and corresponds to a single saddle point
on the energy surface.
The saddle point was found using the minimum mode following method and EMT starting from small random displacement of
the atoms from the Mackay structure, and the steepest descent path then found in both directions
down from the saddle point.
Blue spheres represent atoms in the first shell and
red spheres represent the seven displaced atoms.
}
\label{fig:avalanche}
\end{figure} 
% -------------------------------------------------------------------------------------------------------

%
% -----------------------------------------------  Fig. 2 -----------------------------------------------------
\begin{figure}
\center
\includegraphics[width=.70\textwidth]{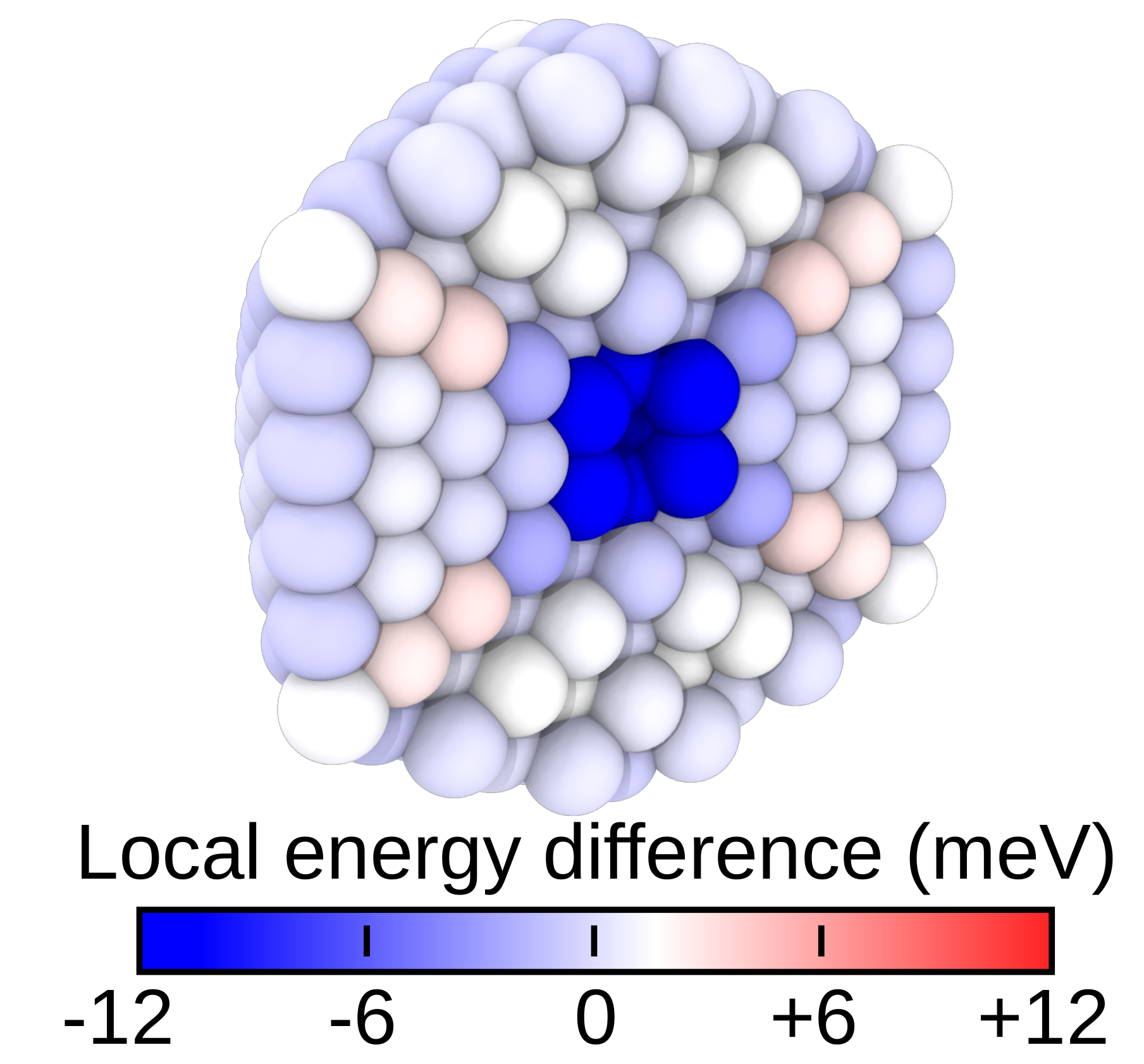}
\caption{
Change in the energy per atom in the 561 Mackay atom icosahedral cluster as the central atom is removed,
calculated using the GAP potential.
Blue and red colors indicate increased and decreased energy per atom, respectively.
The energy lowering is mainly for the innermost atoms, showing how
the 13 atom core of the 561 atom cluster is under pressure from the outer shell atoms. 
}
\label{fig:structures_a}
\end{figure} 
% --------------------------------------------------------------------------------------------------------------

Fig. 2 shows the change in the energy of each atom as the central atom is removed from the 561 atom cluster.
The lowering of the total energy corresponds to 1.6 meV per atom.
The removal of the central atom results in 
lowering of the local strain. The effect is largest in the first shell of 12 atoms and, to a lesser extent in the surface shell. 
In the two subsurface atomic layers 
a handful of atoms increase slightly in energy.

Table 1 summarizes the main characteristics of the 3 and 4 shell icosahedral clusters obtained in this way.
The larger cluster is shown in Fig. 3. The two surface facets containing extra rows of atoms are highlighted.
%
% -----------------------------------------------  Fig. 3 -----------------------------------------------------
\begin{figure}
\center
\includegraphics[width=.83\textwidth]{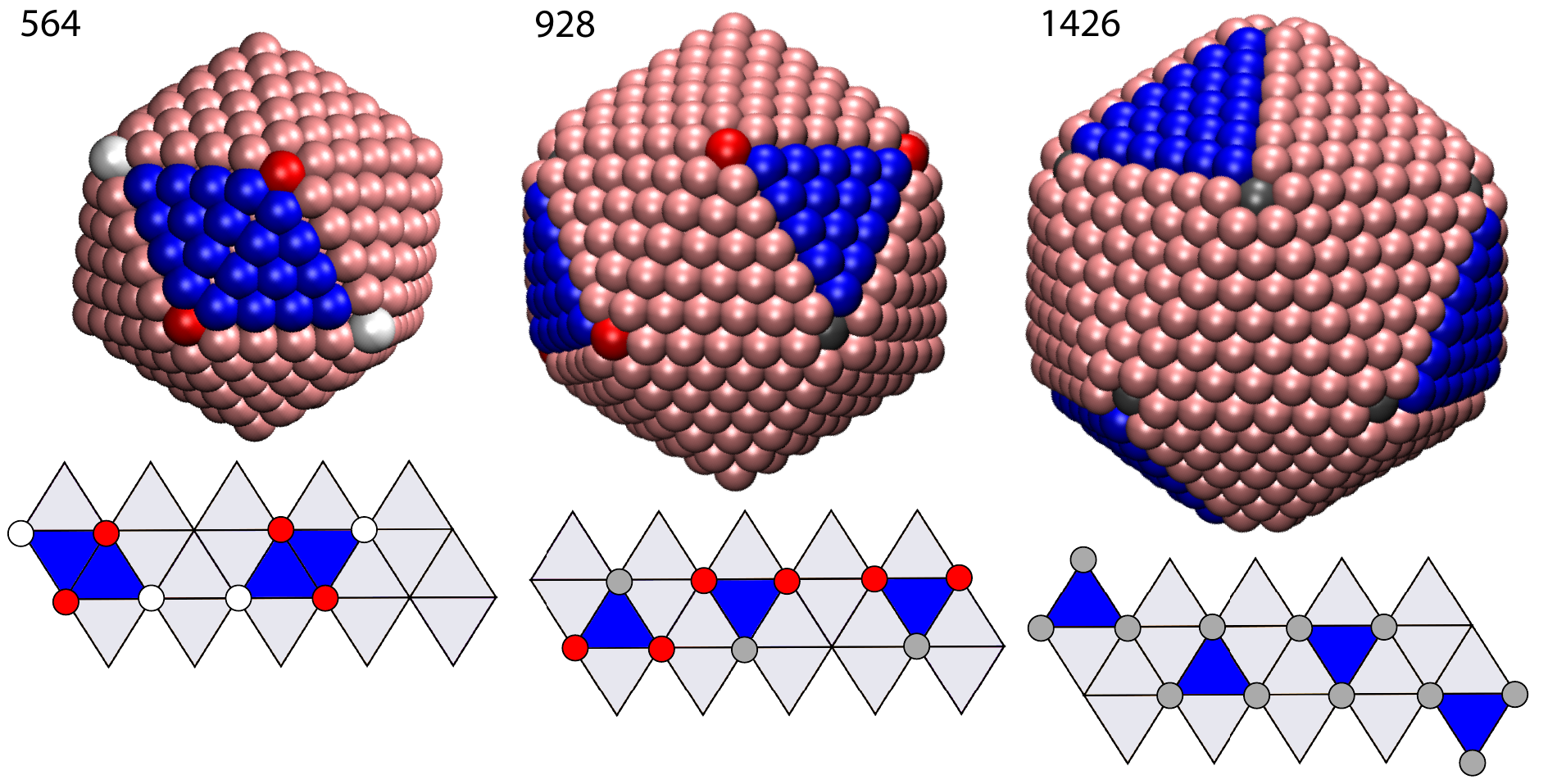}
\caption{
Optimal icosahedral structure of clusters with 564, 927, and 1426 atoms formed by adding  
rows of atoms to the Mackay icosahedra and creating a central vacancy, as listed in Table 1.
Facets where additional atoms have been inserted are shown in blue.
Vertex atoms at anomalous sites and belonging to the surface shell are shown in white, while
subsurface atoms are shown in gray (seen when vertex atoms are missing). 
Atoms that have moved out from a subsurface shell to a location close to the center of a hexagonal ring are shown in red.
The three types of anomalous vertex sites are shown in Fig.~\ref{fig:vertex}.
}
\label{fig:structures_b}
\end{figure} 
% --------------------------------------------------------------------------------------------------------------

% ----------------------------------------------------------  Table 1 -------------------------------------------------------
\begin{table}
\caption{
Characterization of the optimal structure of 3, 4, 5 and 6 layer Au icosahedra. 
In all cases, an atom has been removed from the central site to create a vacancy and atoms 
have been added to the surface shell so as to find the lowest energy per atom in 
DFT/PBEsol calculations. 
The anomalous vertex sites are defined and shown in Fig.~\ref{fig:vertex}. 
}
\begin{tabular}{l|ccccccccc}
\hline
Number of atoms			&&310	 &&  564   	&& 928    &&1426 \\
Atoms per row				&&3		&& 4		&& 5	      &&  6    \\\hline
Comparison w. nearest Mackay:  &&		&&	         &&        && \\
Addional atoms in cluster           &&1	        && 3           && 5    && 11  \\
Additional atoms in facets		&&6		&& 8		&& 15    && 24   \\
Additional rows	of atoms	         &&2		&& 2		&& 3	    && 4    \\\hline
Vertex sites:    	                          &&           &&		&&	    &&    \\
Capped pentagons			&&4		&&4		&& 3	    && 0    \\
Distorted hexagons			&&4		&&4		&& 0	    && 0    \\
Open hexagons                 	&&4		&&0		&& 3	    && 12   \\ 
Filled hexagons	         		&&0		&&4		&& 6	    && 0   \\ 
\hline
\end{tabular}
\label{tab:changeAtoms}
\end{table}

% ---------------------------------------------------------------------------------------------------------------

The DFT calculated energy per atom in these clusters compared with that of the FCC crystal, 
{\it i.e.}, the excess energy per atom, is given in Table 2 and compared with Mackay icosahedral structures 
where the central atom has been removed but no extra atoms added to the surface shell.  
The addition of atoms to the surface shell is found to lower the energy of the cluster with four shells 
(containing 560 vs. 564 atoms) 
by {\it ca.} 10 meV per atom. 
Similar energy lowering due to the surface reconstruction and addition of surface atoms is found with the 
two density functionals, PBEsol and PBE, and with the GAP potential. 

% ----------------------------------- Table 2  ----------------------------------------------------------------
\begin{table}
\ 
\vskip 0.5 true cm
\caption{
Excess energy per atom in meV of the Mackay icosahedral clusters with a central vacancy (MV) and
the optimized cluster structures (Opt) as compared with the calculated cohesive energy of the FCC crystal.
The energy of the clusters as well as that of the FCC crystal has been minimized using
DFT with either the PBEsol or PBE functional approximation, or with the GAP potential function.
}
\begin{tabular}{lcccccccccccccccc}
\\ \hline
                 &&     MV        &       Opt         &&       MV            &     Opt       &&     MV           & Opt           &&          MV          & Opt \\ 
\# atoms	&&      308           &      310      &&       560            &     564        &&      922         &  928           &&       1414      &  1426    \\\hline
PBE	        &&      218           &      209       &&      165            &      155        &&     125         &   115           &&         103      &     93   \\
PBEsol	&&     298           &        292     &&       231            &      221        &&     179         &    171         &&          148      &   140  \\
GAP         &&     286           &         268    &&       215           &       210        &&     181          &   176          &&           156     &   151  \\\hline
\end{tabular}
\label{tab:energiesIcoClosed}
\end{table}
% ----------------------------------------------------------------------------------------------------------------

The vertex sites adjacent to facets containing an extra row of atoms are quite different from the vertex sites of a Mackay icosahedron.
In the 564 atom cluster, such sites have the 
vertex atom sitting on a distorted hexagon instead of a pentagon. 
Vertex sites adjacent to two facets with an extra row are vacant as the vertex atom has become a surface facet atom but
a subsurface atom has been displaced outwards into the surface shell to a location near the center of the hexagon, see Figs. 4 and 5. 
The distance between this displaced atom and the underlying atom is large, 4.4~\AA, 
compared with the near neighbor distance of 2.93~\AA\ in the Au crystal.
Anomalous vertex sites in the 310 atom cluster resemble those of the 564 atom cluster, except that the subsurface atoms are not 
displaced outwards at the vacant vertex sites.  Fig. 3. shows the atomic structure of the three types of anomalous vertex sites.
The intermediate size clusters, 564 and 928, are special in that they have vertex Au atoms that are only sixfold coordinated.

% -----------------------------------------------  Fig. 4 -----------------------------------------------------
\begin{figure}
\center
\includegraphics[width=.83\textwidth]{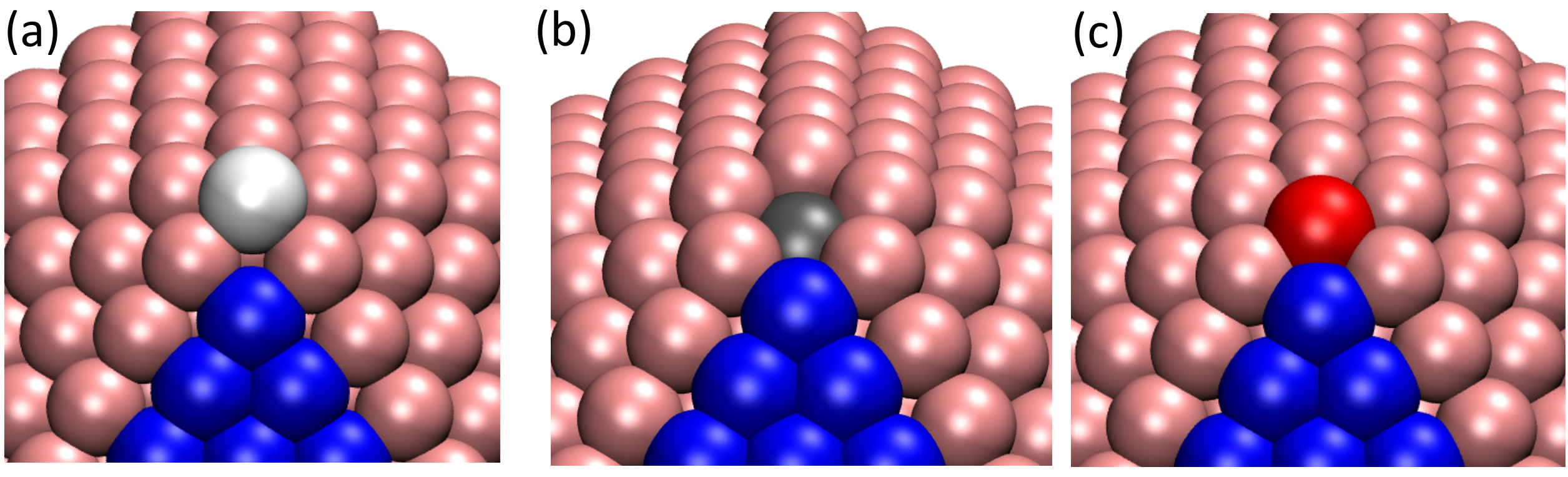}
\caption{
The three anomalous vertex sites formed at facets (shown in blue) that include an extra row of atoms
as compared with the Mackay icosahedra.
(a) Distorted hexagon: The vertex atom (white) is present and sits on a distorted hexagon instead of a pentagon.
(b) Open hexagon: The vertex atom is missing and the underlying subsurface atom (grey) remains in its site in the subsurface shell.
(c) Filled hexagon: The vertex atom is missing and the underlying subsurface atom (red) moves outwards close to the center of a hexagon of atoms,
becoming only sixfold coordinated as the distance to its underlying neighbor becomes 4.4 \AA.  
}
\label{fig:vertex}
\end{figure}
% -------------------------------------------------------------------------------------------------------------

% -----------------------------------------------  Fig. 5 -----------------------------------------------------
\begin{figure}
\center
\includegraphics[width=.7\textwidth]{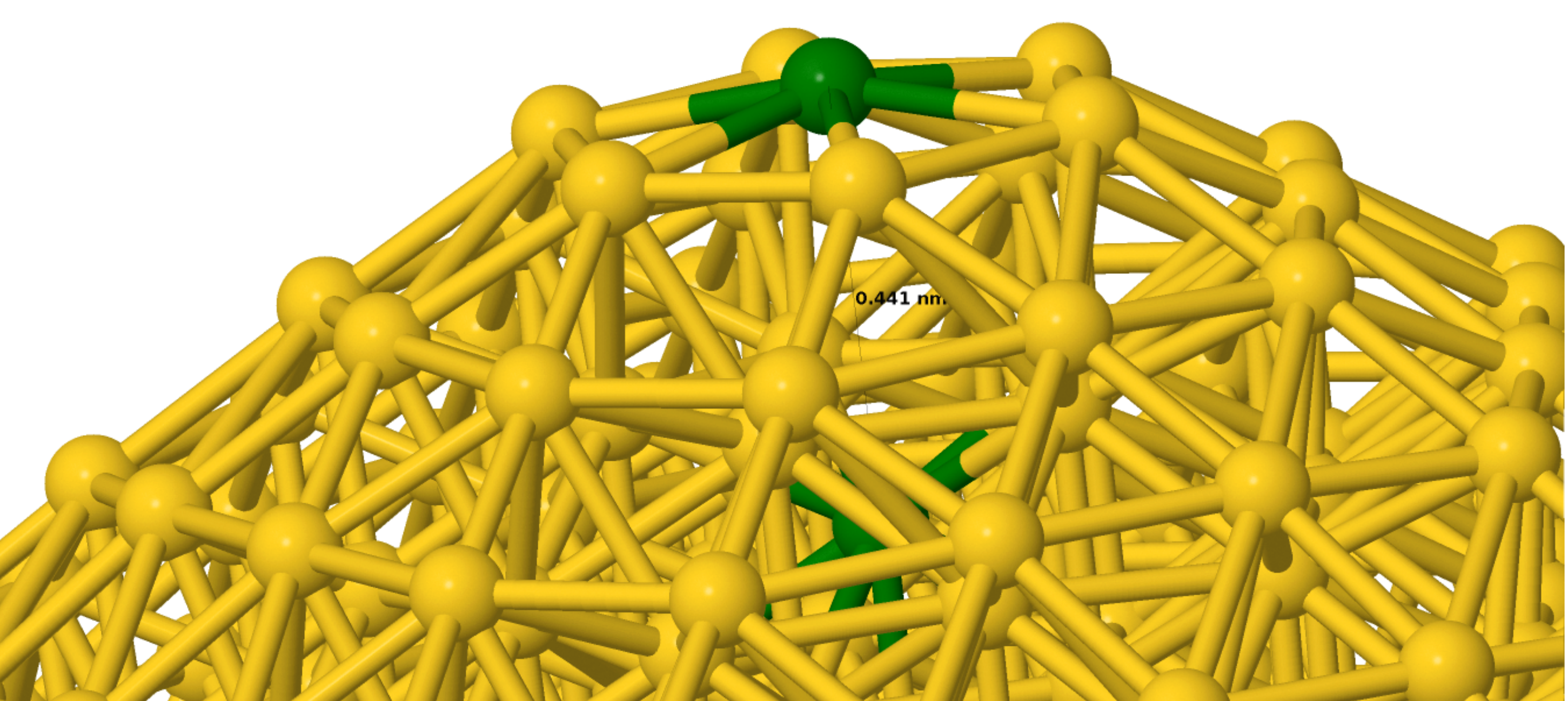}
\caption{ 
The filled hexagon site where the vertex atom of the Mackay icosahedron has been removed and the underlying subsurface atom moves outwards close to the center of a hexagon of atoms. This atom and its nearest neighbor in an inner shell are shown in green. 
The distance between the two is large, 4.4 \AA.
}
\label{fig:bond_a}
\end{figure} 
% ----------------------------------------------------------------------------------------------------------------

Global optimization simulations of the larger, 5 and 6 shell, icosahedral clusters starting from the Mackay icosahedra is more challenging because
of the large number of atoms. 
But, we have generated structures for these clusters by assuming similar motifs as for the smaller clusters, 
{\it i.e.} the removal of the central atom and introduction of extra surface atoms to form additional rows in some of the facets. 
Subsequent GOUST optimizations were carried out using EMT, 
followed by local minimization using GAP or DFT.
The optimal cluster structures found in this way are compared with Mackay icosahedra in Table 1 and illustrated in Fig. 3.
The lowering of the energy per atom as compared with the corresponding Mackay icosahedron with a central vacancy is 8 meV per atom 
using DFT/PBEsol and 10 meV per atom using DFT/PBE for both 5 and 6 shell clusters (see Table 2). 
The GAP potential gives somewhat smaller change in the energy but the same trend.
The largest cluster, containing 1426 atoms, has regained certain symmetry, as illustrated in Fig. 3. 
All 12 vertex sites are vacant there. 

% -----------------------------------------------  Fig. 6 -----------------------------------------------------
\begin{figure}
\center
\includegraphics[width=.7\textwidth]{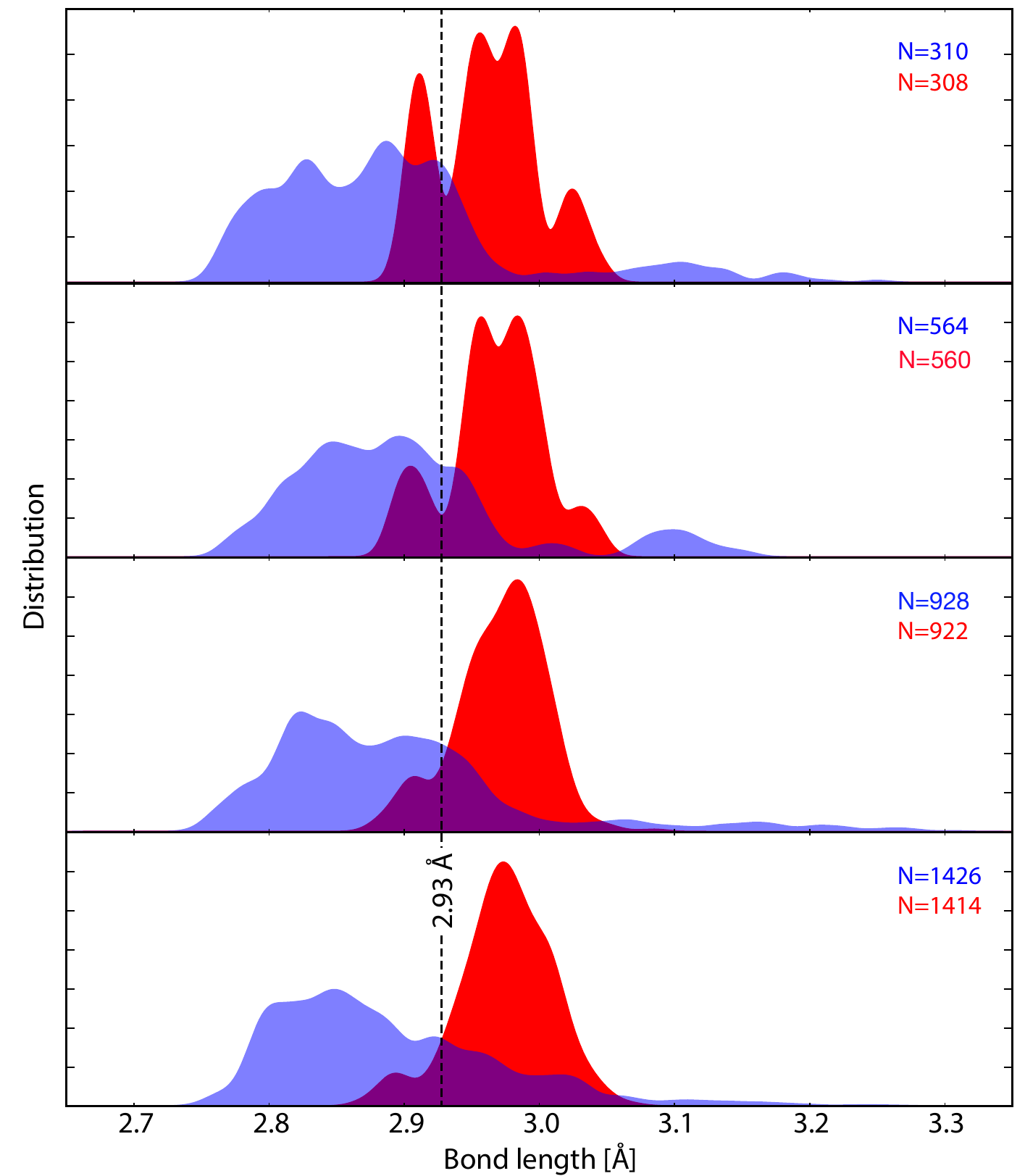}
\caption{ 
Distribution of nearest neighbor distances between surface atoms in the optimized icosahedral clusters (blue) and in 
the Mackay icosahedra with a central vacancy (red). 
The energy has been minimized using DFT/PBEsol. 
The dashed black line marks the calculated near-neighbor distance in the Au crystal at the same level of theory.
The optimized icosahedral structures have a broader distribution and a smaller average distance between neighbors than the Mackay icosahedra.
}
\label{fig:bond_b}
\end{figure} 
% ----------------------------------------------------------------------------------------------------------------

The distance between nearest neighbors in the surface shells of the optimized structures varies greatly, 
much more than in Mackay icosahedra, as illustrated in Fig. 5. 
Nearest neighbor distances ranging from less than 2.8~\AA\ to larger than 3~\AA\ can be found in the optimized structures.
On average, the nearest-neighbor distances are smaller in the optimized structures 
as would be expected from the increased number of atoms in the surface shell.

% ------------------------------------------------------------------------------------------------------------------------------------

\section{Discussion}

The optimal structures presented here for icosahedral Au clusters with 3, 4, 5 and 6 shells of atoms are significantly lower in
energy than the Mackay icosahedra.
The `magic numbers' turn out to be 310, 564 , 928 and 1426 instead of the Mackay magic numbers of 309, 561, 923 and 1415.
Since knowledge of the atomic structure is the basis for
understanding the various properties of the clusters, such as catalysis, this is an important finding.

The fact that the surface shell 
includes additional atoms as compared with the Mackay structure is consistent with the experimental findings for smaller
gold clusters with 55 to 64 atoms\cite{Huang08} and the known surface reconstruction of Au crystal surfaces.\cite{vanHove81}
The tendency of Au clusters and the gold crystal surfaces to accept additional atoms in the outermost layers can be related to relativistic effects.\cite{Huang08}

The unusual atomic ordering at some of the vertex sites and the large variation in nearest-neighbor distances is likely to have
significant effect on the catalytic activity of these clusters.  
The anomalous vertex sites formed where one or two of the adjacent facets include an extra row of atoms are particularly interesting.
Previously, sites with a missing vertex atom and an open hexagonal ring had been
found from DFT calculations of 55 atom Pt clusters.\cite{Apra04} There, these sites were referred to as `rosettes'.  
Here, we also observe the possibility that an underlying atom moves outward to a location that is near the center of the hexagonal ring.
This is the case for clusters of intermediate size, 564 and 928.
The atom near the center of a hexagon is only sixfold coordinated, since the distance to the nearest inner shell neighbor is 4.4\AA. 
Such low-coordinated atoms are likely to provide higher binding energy for adatoms and admolecules, and also have significant structural flexibility 
since a displacement toward the center does not involve large increase in energy.
Such structural flexibility could be of importance in chemical reactions.

The saddle point searches used in the optimization method applied here readily revealed the formation of a vacancy at 
the center of the clusters. 
This lowers the energy per atom because the center is under compressive strain in large icosahedral clusters. 
Such vacancy formation had previously been predicted for icosahedral Pt clusters based on tight binding 
calculations\cite{Mottet97} and is likely a common feature in icosahedral metal clusters.

% ------------------------------------------------------------------------------------------------------------------------------------

\section{Acknowledgements}
This work was supported by the Icelandic Research Fund and
the Academy of Finland under grants number 329483 (J.K. and M.A.C.) and 330488 (M.A.C.). 
Computational resources from CSC - IT Center for Science (NOBLE Grand Challenge project) and Aalto University's Science IT Project are gratefully acknowledged as well as
the Icelandic Research High Performance Computing facility.
We thank Anna L. Garden for helpful discussions.

\vfill\eject
\newpage

% ----------------------------------------------------------------------------------------------------------------------------------------------------

% --------------------------------------------------

\begin{thebibliography}{99}

% --------

\bibitem{Haruta97} M. Haruta, 
{\it Catal. Today}, 1997, {\bf 36}, 153.
%Ã166.

\bibitem{Valden98} M. Valden, X. Lai and D. W. Goodman,  
{\it Science}, 1998, {\bf 281}, 1647.
%Ã1650.

\bibitem{Corma13}  A. Corma {\it et al.}, 
%Patricia ConcepciÂn, Mercedes Boronat, Maria J. Sabater, Javier Navas, Miguel JosÂ Yacaman, Eduardo Larios, Alvaro Posadas, M. Arturo LÂpez-Quintela, David Buceta, Ernest Mendoza, Gemma Guilera and Alvaro Mayoral, Ã
%Exceptional oxidation activity with size-controlled supported gold clusters of low atomicity.
 {\it  Nat. Chem.}, 2013, {\bf 5}, 775. 

\bibitem{Saint-Lager13}  M-C. Saint-Lager, I. Laoufi and A. Bailly,
%Operando atomic structure and active sites of TiO2(110)-supported gold nanoparticles during carbon monoxide oxidation
%  show a nice peak in catalytic activity for clusters that are between 2 and 3 nm
{\it Faraday Discuss.}, 2013, {\bf 162}, 179.

\bibitem{Maier01} S. A. Maier, M. L. Brongersma, P. G. Kiki, S. Meltzer, A. A. G. Requicha and H. A. Atwater, 
{\it Adv. Mater.}, 2001, {\bf 13}, 1501.
%Ã1505.

\bibitem{Ghosh07}  S.K. Ghosh and T. Pal, 
%Intercoupling Coupling Effect on the Surface Plasmon Resonance of Gold Nanoparticles: From Theory to Applications. 
{\it Chemical Reviews.}, 2007, {\bf 107}, 4797.

\bibitem{Saha12} K. Saha, S. S. Agasti, C. Kim, X. Li and V. M. Rotello, 
{\it Chem. Rev.}, 2012, {\bf 112}, 2739.
%  biosensors

% -----

\bibitem{Brodersen11} S. H. Brodersen, U. Gr{\o}nbjerg, Britt Hvolb{\ae}k and Jakob Schi{\o}tz,
% Understanding the catalytic activity of gold nanoparticles through multi-scale simulations. 
%We investigate how the chemical reactivity of gold nanoparticles depends on the cluster size and shape using a combination of simulation techniques at different length scales, enabling us to model at the atomic level the shapes of clusters in the size range relevant for catalysis. The detailed atomic configura- tion of a nanoparticle with a given number of atoms is calculated by first finding overall cluster shapes with low energy and approximately the right size, and then using Metropolis Monte Carlo simulations to identify the detailed atomic configuration. The equilibrium number of low-coordinated active sites is found, and their reactivities are extracted from models based on Density Functional Theory calculations. This enables us to determine the chemical activity of clusters in the same range of particle sizes that is accessible experimentally. The variation of reactivity with particle size is in excellent agreement with experiments, and we conclude that the experimentally observed trends are mostly explained by the high reactivity of under-coordinated corner atoms on the gold clusters. Other effects, such as the effect of the substrate, may influence the reactivities significantly, but the presence of under-coordinated atoms is sufficient to explain the overall trend.
{\it J. Catal.}, 2011, {\bf 284}, 34.

\bibitem{Assa09}  B. Assadollahzadeh and P. Schwerdtfeger, 
{\it J. Chem. Phys.}, 2009, {\bf 131}, 064306.

\bibitem{Michaelian99} K. Michaelian, N. Rendon and I.L. Garzon,
% Structure and Energetics of Ni, Ag, and Au Nanoclusters. 
{\it Phys. Rev. B}, 1999, {\bf 60}, 2000.

\bibitem{Li03} J. Li, X. Li, H-J. Zhai and L-S. Wang,
%Au20: A Tetrahedral Cluster
{\it Science}, 2003, {\bf 299}, 864.

\bibitem{Baletto05}  F. Baletto and R. Ferrando, 
{\it Rev. Mod. Phys.}, 2005, {\bf 77}, 371.
%Structural properties of nanoclusters: Energetic, thermodynamic, and kinetic effects.

\bibitem{Pyykko08}  Pyykk\"o, 
{\it Chem. Soc. Rev.}, 2008, {\bf 37}, 1967.

\bibitem{Huang08}  W. Huang, M. Ji, C.-D. Dong, X. Gu, L.-M. Wang, X. G. Gong and L.-S. Wang,  
%  Relativistic Effects and the Unique Low-Symmetry Structures of Gold Nanoclusters
{\it ACS Nano}, 2008, {\bf 2}, 897.

\bibitem{Wang12a} Z. W. Wang and R. E. Palmer,  
% Experimental Evidence for Fluctuating, Chiral-Type Au55 Clusters by Direct Atomic Imaging
{\it Nano Lett.}, 2012, {\bf 12}, 5510.

% ---------

\bibitem{Honeycutt87} J. D. Honeycutt and H. C. Andersen,
% Molecular dynamics study of melting and freezing of small Lennard-Jones clusters
{\it J. Phys. Chem.}, 1987, {\bf 91}, 4950.

\bibitem{Mackay62}  A. L. Mackay, 
% A dense non-crystallographic packing of equal spheres.
{\it Acta. Cryst.}, 1962, {\bf 15}, 916.

\bibitem{Ino69} S. Ino,
%  Stability of multiply-twinned particles.
{\it Journal of the Physical Society of Japan},  1969, {\bf 27},  941.

\bibitem{Knight84} W. D. Knight, K. Clemenger, W. A. de Heer, W. A. Saunders, M. Y. Chou, and M. L. Cohen, 
{\it Phys. Rev. Lett.}, 1984, {\bf 52}, 2141. 

\bibitem{Wrigge02} G. Wrigge, M. Astruc Hoffmann, and B. v. Issendorff, 
%  Photoelectron spectroscopy of sodium clusters: Direct observation of the electronic shell structure
{\it Phys. Rev. A}, 2002, {\bf 65}, 063201.

\bibitem{Larsen11} A.H. Larsen, J. Kleis, K.S. Thygesen, J.K. N{\o}rskov and K.W. Jacobsen,
%Electronic shell structure and chemisorption on gold nanoparticles
{\it Phys. Rev. B}, 2011, {\bf 84}, 245429.


%  Au55
\bibitem{Garzon96}  I.L. Garz\'on and A. Posada-Amarillas, 
%  Structural and vibrational analysis of amorphous Au 55 clusters
%      use common neighbor analysis and refer to Daniel's paper
{\it Phys. Rev. B}, 1996,  {\bf 54}, 11796.

\bibitem{Garzon98} I.L. Garz\'on, K. Michaelian, M.R. Beltr\'an, A. Posada-Amarillas, P. Ordej\'on, E. Artacho, D. S\'anchez-Portal and J. M. Soler, 
{\it Phys. Rev. Lett.}, 1998,  {\bf 81}, 1600.
%  Baletto review states: The two rosettes of the double rosette are at nearby vertices, giving a close resemblance to the global minimum of Au55 found by Garzn, Michaelian, et al.

%Lo?ez-Lozano,X.;Pe?ez,L.A.;Garzo?,I.L.
% Enantiospecific Adsorption of Chiral Molecules on Chiral Gold Clusters.
%{\it Phys. Rev. Lett.}, 2006,  {\bf 97}, 233401.


% ---  surf reconstr.

\bibitem{vanHove81} M.A. van Hove, R.J. Koestner, P.C. Stair, J.P. Biberian, L.L. Kesmodel, I. Bartos and G.A. Somorjai, 
% The Surface Reconstructions of the (100) Crystal Faces of Iridium, Platinum and Gold .1. Experimental-Observations and Possible Structural Models. 
{\it Surf. Sci.}, 1981, {\bf 103},  189.

\bibitem{Takeuchi91} N. Takeuchi, C.T. Chan and K.M. Ho, 
%Reconstruction of the (100) Surfaces of Au and Ag. 
{\it Phys. Rev. B}, 1991, {\bf 43}, 14363.
%Ð 14370.


% --  SC-STEM on Au clusters

\bibitem{Li08}
%Li, Z. Y.; Young, N. P.; Di Vece, M.; Palomba, S.; Palmer, R. E.; Bleloch, A. L.; Curley, B. C.; Johnston, R. L.; Jiang, J.; Yuan, J. 
%Three-dimensional atomic-scale structure of size-selected gold nanoclusters. 
%Nature 2008, 451, 46–48.
Z.Y. Li, N.P. Young, M. Di Vece, S. Palomba, R.E. Palmer, A.L. Bleloch, B.C. Curley, R.L. Johnston, J. Jiang and J. Yuan, 
% Three-dimensional atomic-scale structure of size-selected gold nanoclusters.
%    Au309
{\it Nature}, 2008, {\bf 451},  46.

\bibitem{Curley07} B. C. Curley, R. L. Johnston, N. P. Young, Z. Y. Li, M. Di Vece, R. E. Palmer and A. L. Bleloch, 
% Combining Theory and Experiment to Characterize the Atomic Structures of Surface-Deposited Au309 Clusters 
{\it J. Phys Chem. C }, 2007, {\bf 111}, 17846.
% ««Interestingly, for Au309 itself, a distorted icosahedral structure was found to be more stable than the regular icosahedron. As shown in Figure 7, the nature of the distortion in Au309 is a puckering deformation, corresponding to the depression of some of the icosahedral vertex atoms, leading to ÒrosetteÓ formation, as has been observed previously for smaller clusters, such as Pt55 [Apra04].
%  the complete 309-atom Ih cluster is a magic structure or globe minimum, whose energy can be further reduced by a rosette distortion on the cluster surface, as previously found for Au309.
%     they seem to think that the vacancy only forms at elevated temperature

\bibitem{Wang12b} Z. W. Wang and R. E. Palmer, 
% Determination of the Ground-State Atomic Structures of Size-Selected Au Nanoclusters by Electron-Beam-Induced Transformation.
{\it Phys. Rev. Lett.}, 2012, {\bf 108}, 245502.

%   larger clusters:

\bibitem{Plant14} S.R. Plant, L. Cao and R.E. Palmer, 
% Atomic Structure Control of Size- Selected Gold Nanoclusters during Formation.
{\it J. Am. Chem. Soc.} 2014, {\bf 136}, 7559.

\bibitem{Wells15}  D. M. Wells, G. Rossi, R. Ferrando and R. E. Palmer, 
{\it Nanoscale}, 2015, {\bf 7}, 6498.

\bibitem{Baletto02}  F. Baletto, R. Ferrando, A.  Fortunelli,  F. Montalenti and C. Mottet, 
% Crossover among structural motifs in transition and noble-metal clusters.
{\it J. Chem. Phys.}, 2002, {\bf 116}, 3856.

\bibitem{Grochola07}  G. Grochola, I. K. Snook and S. P. Russo,  
{\it J. Chem. Phys.}, 2007, {\bf 127}, 224704.

\bibitem{Cleveland97}  C. L. Cleveland, U. Landman, M. N. Shafigullin, P. W. Stephens and R. L. Whetten, 
{\it Z. Phys. D}, 1997, {\bf 40}, 503.
%Ð508.

\bibitem{Uppenbrink92}  J. Uppenbrink, D. J. Wales,  
{\it J. Chem. Phys.}, 1992, {\bf 96}, 8520.
%Ð8534.

\bibitem{Negreiros07}  F. R. Negreiros, E. A. Soares and V. E. de Carvalho,  
{\it Phys. Rev. B}, 2007, {\bf 76}, 205429.

\bibitem{Bao09}  K. Bao, S. Goedecker, K. Koga, F. Lancon and A. Neelov, 
{\it Phys. Rev. B}, 2009, {\bf 79}, 041405.

\bibitem{Garden18} A. L. Garden, A. Pedersen and H. J\'onsson,
%  Reassignment of `magic numbers' for Au clusters of decahedral and FCC structural motifs,
{\it Nanoscale}, 2018, {\bf 10}, 5124.



\bibitem{Barnard09}  A. S. Barnard, N. P. Young, A. Kirkland, M. van Huis and H. Xu, 
% Nanogold: a quantitative phase map.
{\it ACS Nano}, 2009, {\bf 3}, 1431.
%Ð1436.

\bibitem{Rahm17}  J. M Rahm and P. Erhart,  
{\it Nano Lett.}, 2017, {\bf 17}, 5775.
% Beyond Magic Numbers: Atomic Scale Equilibrium Nanoparticle Shapes for Any Size
%In the pursuit of complete control over morphology in nanoparticle synthesis, knowledge of the thermodynamic equilibrium shapes is a key ingredient. While approaches exist to determine the equilibrium shape in the large size limit (≲10-20 nm) as well as for very small particles (≲2 nm), the experimentally increasingly important intermediate size regime has largely remained elusive. Here, we present an algorithm, based on atomistic simulations in a constrained thermodynamic ensemble, that efficiently predicts equilibrium shapes for any number of atoms in the range from a few tens to many thousands of atoms. We apply the algorithm to Cu, Ag, Au and Pd particles with diameters between approximately 1 and 7 nm and reveal an energy landscape that is more intricate than previously suggested. The thus obtained particle type distributions demonstrate that the transition from icosahedral particles to decahedral and further into truncated octahedral particles occurs only very gradually, which has implications for the interpretation of experimental data. The approach presented here is extensible to alloys and can in principle also be adapted to represent different chemical environments.

\bibitem{Marks84} L.D. Marks, 
% Surface-Structure and Energetics of Multiply Twinned Particles.
{\it Philos. Mag. A}, 1984, {\bf 49},  81.

\bibitem{Marks94} L. D. Marks, 
% Experimental studies of small particle structures
{\it Rep. Prog. Phys.}, 1994,  {\bf 57}, 603. 


\bibitem{Koga04}  K. Koga, T. Ikeshoji and  K Sugawara,  
% Size- and Temperature-Dependent Struc- tural Transitions in Gold Nanoparticles. 
{\it Phys. Rev. Lett.}, 2004, {\bf 92}, 115507.


\bibitem{Doye95}  Doye, J. P. K., Wales, D. 
{\it J. Chem. Phys. Lett.}, 1995, {\bf 247}, 339.




\bibitem{Chen11} F. Chen, Z.Y. Li and R.L. Johnston, 
% Surface reconstruction precursor to melting in Au309 clusters
%     from abstract: (111) faceted icosahedral gold cluster, which form a liquid patch due to surface vacancy.
%  abstract:
%  The melting of gold cluster is one of essential properties of nanoparticles and re- visited to clarify the role played by the surface facets in the melting transition by molecular dynamics simulations. The occurrence of elaborate surface reconstruc- tion is observed using many-body Gupta potential as energetic model for 309-atom (2.6 nm) decahedral, cuboctahedral and icosahedral gold clusters. Our results reveal for the first time a surface reconstruction as precursor to the melting transitions. The surface reconstruction lead to an enhanced melting temperature for (100) faceted decahedral and cuboctahedral cluster than (111) faceted icosahedral gold cluster, which form a liquid patch due to surface vacancy. 
{\it AIP Advances },  2011, {\bf 1}, 032105.

\bibitem{Li15} H. Li, L. Li, A. Pedersen, Y. Gao, N. Khetrapal, H. J\'onsson and X.C. Zeng,  
{\it Nano Letters}, 2015, {\bf 15},  682.

% HREM studies, showing prevalence of icosahedra, formation of Ih clusters is governed by kinetic rather than thermodynamic factors.
\bibitem{Buffat91} P.-A. Buffat, M. Fl\"ueli, R. Spycher, P. Stadelmann and J.-P. Borel, 
%  Crystallographic Structure of Small Gold Particles studied by High-resolution Electron Microscopy
{\it Faraday Discuss.}, 1991, {\bf 92}, 173.

\bibitem{Kuo02} K.H. Kuo, 
% Mackay, anti-Mackay, double-Mackay, pseudo-Mackay, and related icosahedral shell clusters.
{\it Structural Chemistry}, 2002, {\bf 13},  221.

\bibitem{Mottet97}  C. Mottet, G. Tr\'eglia, and B. Legrand,  
% New magic numbers in metallic clusters: an unexpected metal dependence.
{\it Surf. Sci.}, 1997, {\bf 383}, L719.
%ÐL727.


% ------------------------  
%     Methods

\bibitem{Plasencia14} M. Plasencia, A. Pedersen, A. Arnaldsson, J-C. Berthet and H.  J\'onsson,
% Geothermal model calibration using a global minimization algorithm based on finding saddle points as well as minima of the objective function
{\it Computers and Geosciences}, 2014, {\bf 65}, 110.

\bibitem{Henkelman99} G. Henkelma and H. J\'onsson,
% A dimer method for finding saddle points on high dimensional potential surfaces using only first derivatives.
{\it J. Chem. Phys.}, 1999, {\bf 111},  7010.

%\bibitem{Henkelman02}  G. Henkelman, G. J\'ohannesson, and H. J\'onsson, 
%Theoretical Methods in Condensed Phase Chemistry (Kluwer Academic, New York, 2002), Vol. 5, p. 269.

%\bibitem{Olsen04} R. A. Olsen, G. J. Kroes, G. Henkelman, A. Arnaldsson and H. J\'onsson, 
%Comparison of methods for finding saddle points without knowledge of the final states, 
%{\it J. Chem. Phys.}, 2004, {\bf 121}, 9776.

\bibitem{Pedersen11} A. Pedersen, S.F. Hafstein and H. J\'onsson, 
% Efficient Sampling of Saddle Points with the Minimum-Mode Following Method.
{\it SIAM Journal on Scientific Computing},  2011, {\bf 33},  633.

\bibitem{Gutierrez17}
%Improved Minimum Mode Following Method for Finding First Order Saddle Points', 
M.P. Gutirr\'ez, C. Arg\'aez and H. J\'onsson, 
{\it J. Chem. Theory Comput.}, 2017, {\bf13}, 125.

%\bibitem{Pedersen14} A. Pedersen and M. Luiser, 
% Bowl breakout: Escaping the positive region when searching for saddle points. 
%{\it J. Chem. Phys.},  2014, {\bf 141},  024109.

\bibitem{Chill14} S.T. Chill, M. Welborn, R. Terrell, L. Zhang, J-C. Berthet, A. Pedersen, H. J\'onsson and G. Henkelman, 
%  EON: software for long time simulations of atomic scale systems} 
{\it Modelling and Simulation in Materials Science and Engineering},  2014, {\bf 22},  055002.

%\bibitem{ASAP} https://wiki.fysik.dtu.dk/asap. 
%The Rasmussen set of parameter values for Au were used in the EMT potential function. 

\bibitem{Jacobsen96} K.W. Jacobsen, P. Stoltze and J.K. N{\o}rskov, 
% A semi-empirical effective medium theory for metals and alloys.
{\it Surf. Sci.}, 1996, {\bf 366},  394.

\bibitem{GOUST}
%Geothermal model calibration using a global minimization algorithm based on finding saddle points as well as minima of the objective function', 
M. Plasencia, A. Pedersen, A. Arnaldsson, J-C. Berthet and H. J\'onsson, 
Computers and Geosciences, 2014, {\bf  65}, 110.

\bibitem{Perdew96} J.P. Perdew, K. Burke and M. Ernzerhof, 
% Generalized gradient approximation made simple.
{\it Phys. Rev. Letters}, 1996, {\bf 77},  3865.

\bibitem{Perdew08} J.P. Perdew, A. Ruzsinszky, G. Csonka, O. Vydrov, G. Scuseria, L. Constantin, X. Zhou and K. Burke, 
% Restoring the Density-Gradient Expansion for Exchange in Solids and Surfaces. 
%  PBEsol
{\it Phys. Rev. Letters}, 2008, {\bf 100}, 136406.

\bibitem{Blochl94} P.E. Bl\"ochl, 
% Projector augmented-wave method.
{\it Phys. Rev. B}, 1994, {\bf 50},  17953.

\bibitem{Kresse96a} G. Kresse and J. Furthm\"uller, 
% Efficiency of ab-initio total energy calculations for metals and semiconductors using a plane-wave basis set.
{\it Comput. Mat. Sci.}, 1996, {\bf 6}, 15.

\bibitem{Kresse96b} G. Kresse and J. Furthm\"uller,
% Efficient iterative schemes for ab initio total-energy calculations using a plane-wave basis set. 
{\it Phys. Rev. B}, 1996, {\bf 54},  11169.

\bibitem{GAPa}
%Gaussian approximation potentials: The accuracy of quantum mechanics, without the electrons. 
A.P. Bart\'ok, M.C. Payne, R. Kondor, G. Cs\'anyi. 
{\it Phys. Rev. Lett.}, 2010, {\bf 104}, 136403.

\bibitem{GAPb}
%Machine Learning Interatomic Potentials as Emerging Tools for Materials Science. 
V. L. Deringer, M.A. Caro, G. Cs\'anyi,
{\it Adv. Mater.}, 2019, {\bf 31}, 1902765.

%\bibitem{TurboGAP} http://turbogap.fi 

\bibitem{Deringer21}
V.L. Deringer, A.P. Bart\'ok, N. Bernstein, D.M. Wilkins, M. Ceriotti, and G. Cs\'anyi,
Gaussian Process Regression for Materials and Molecules. 
{\it Chem. Rev.}, 2021, {\bf 121}, 10073.

\bibitem{Kloppenburg22}
J. Kloppenburg. 
GAP interatomic potential for gold. 
Zenodo (2022). DOI: 10.5281/zenodo.6302852.

\bibitem{Bartok13}
A.P. Bartók, R. Kondor, and G. Cs{\'a}nyi. 
On representing chemical environments. 
{\it Phys. Rev. B}, 2013, {\bf 87}, 184115.

\bibitem{Caro19} M.A. Caro. 
Optimizing many-body atomic descriptors for enhanced computational performance of machine learning based interatomic potentials. 
{\it Phys. Rev. B}, 2019, {\bf 100}, 024112.

% -----------------
%      Discussion

\bibitem{Apra04} E. Apr\'a, F. Baletto, R. Ferrando and A Fortunelli, 
% Amorphization Mechanism of Icosahedral Metal Nanoclusters.
{\it Phys. Rev. Letters},  2004, {\bf 93},  065502.
%   Baletto review: rosette structures, originating from the disordering of one or two vertices in icosahedra of 55 atoms, are considerably lower in energy than the icosa- hedron itself. This result would support the preference of Pt for low-symmetry structures at icosahedral magic numbers.

\bibitem{Hendy02} S.C. Hendy and J.P.K. Doye, 
% Surface-reconstructed icosahedral structures for lead clusters
{\it Phys. Rev. Letters}, 2002,  {\bf 66},  235402.




\end{thebibliography}
\end{document}